\documentclass[10pt,leqno,titlepage]{amsart}
\usepackage[foot]{amsaddr}
\usepackage{amsmath}
\usepackage{graphicx,psfrag,epsf}
\usepackage{enumerate}
\usepackage{natbib}
\usepackage{url} %

\usepackage{mathtools, empheq}
\usepackage{amssymb, setspace}
\usepackage{bm, bbm}
\usepackage{graphicx, tabularx, array}
\usepackage[svgnames,table]{xcolor}
\usepackage{enumitem}
\usepackage{multirow}
\usepackage{hyperref}
\usepackage{geometry}
\usepackage{makecell}
\usepackage{booktabs}
\usepackage{amsmath}
\usepackage{array}
\usepackage{algorithm}
\usepackage{algorithmic}

\usepackage{longtable}
\usepackage{threeparttable} %

\usepackage{colortbl} %

\newcolumntype{C}[1]{>{\centering\arraybackslash}m{#1}}

\newcommand{\1}{\mathbbm{1}}

\numberwithin{equation}{section}

\usepackage{amssymb,amsthm,amsmath}

\usepackage{etoolbox}
\AtBeginDocument{%
    \let\orignewpage\newpage 
    \renewcommand\newpage{}
    \patchcmd{\clearpage}{\newpage}{\orignewpage}{}{}}

\begin{document}
\bibliographystyle{plainnat}

\def\spacingset#1{\renewcommand{\baselinestretch}%
{#1}\small\normalsize} \spacingset{1}

\newlist{steps}{enumerate}{1}
\setlist[steps, 1]{label = Step \arabic*:}

\title[AUD Study Protocol]{Protocol for an Observational Study on the Effects of Paternal Alcohol Use Disorder on Children's Later Life Outcomes}

\author{William Bekerman$^{1}$} \email{bekerman@wharton.upenn.edu}
\author{Marina Bogomolov$^{2}$}
\email{marinabo@technion.ac.il}
\author{Ruth Heller$^{3}$}
\email{ruheller@tauex.tau.ac.il}
\author{Matthew Spivey$^{1}$}
\email{spiveym@wharton.upenn.edu}
\author{Kevin G. Lynch$^{4}$}
\email{lynch3@pennmedicine.upenn.edu}
\author{David W. Oslin$^{4}$}
\email{oslin@upenn.edu}
\author{Dylan S. Small$^{1}$}
\email{dsmall@wharton.upenn.edu}

\dedicatory{$^{1}$Department of Statistics and Data Science, University of Pennsylvania, Philadelphia, PA, USA\\$^{2}$Data and Decision Sciences, Technion - Israel Institute of Technology, Haifa, Israel\\$^{3}$Department of Statistics and Operations Research, Tel-Aviv University, Tel-Aviv, Israel\\$^{4}$Department of Psychiatry, University of Pennsylvania, Philadelphia, PA, USA}

\begin{abstract}
The harmful effects of growing up with a parent with an alcohol use disorder have been closely examined in children and adolescents, and are reported to include mental and physical health problems, interpersonal difficulties, and a worsened risk of future substance use disorders. However, few studies have investigated how these impacts evolve into later life adulthood, leaving the ensuing long-term effects of interest. In this article, we provide the protocol for our observational study of the long-term consequences of growing up with a father who had an alcohol use disorder. We will use data from the Wisconsin Longitudinal Study to examine impacts on long-term economic success, interpersonal relationships, physical, and mental health. To reinforce our findings, we will conduct this investigation on two discrete subpopulations of individuals in our study, allowing us to analyze the replicability of our conclusions. We introduce a novel statistical design, called data turnover, to carry out this analysis. Data turnover allows a single group of statisticians and domain experts to work together to assess the strength of evidence gathered across multiple data splits, while incorporating both qualitative and quantitative findings from data exploration. We delineate our analysis plan using this new method and conclude with a brief discussion of some additional considerations for our study.
\end{abstract}

\maketitle

\spacingset{1.5}

\section{Background and Motivation}

Alcohol use disorder (AUD), also known colloquially as alcohol dependence or alcoholism, has long been reported to run in families and numerous genetic factors have been identified as contributing to a greater risk of developing the disorder (\cite{edenberg2013genetics}). Along with this predisposition to future alcohol use disorders, growing up with a parent with AUD can adversely affect child and adolescent development, as well as their mental and physical health. Problematic drinking by parents can contribute to inconsistency or unpredictability in parenting behaviors and may also induce higher levels of marital conflict, a destabilizing influence on families associated with higher levels of child and adolescent alcohol use and aggression. It can also lead parents to model ineffective coping strategies and other problem behaviors (\cite{windle1996effect}). Moreover, children and adolescents of parents with AUD often have a less healthy lifestyle and more mental health problems compared to their peers (\cite{serec2012health}) and are more frequently affected by depression, anxiety, suicidal ideation, substance abuse, and interpersonal difficulties (\cite{park2015systematic}). 

Unfortunately, these damaging effects are not uncommon and approximately 10.5\% of children in the United States under 18 years of age live with a parent who has an alcohol use disorder (\cite{lipari2017children}). Yet despite its prevalence, most studies on the impacts of growing up with a parent with AUD focus on the drawbacks experienced during childhood and adolescence (see \cite{rossow2016parental} for a review), with fewer studying effects among young adults (e.g., \cite{christoffersen2003long,klostermann2011coping}), and most research on adult outcomes limited to alcohol abuse or psychiatric problems (e.g., \cite{anda2002adverse}). Longitudinal studies such as \cite{beardslee1986exposure} present a unique opportunity to follow these children and examine how these issues evolve throughout their lives. We are interested in how various aspects of the later lives of these persons, particularly related to their health and wellness, are influenced by having grown up with a father who experienced alcohol use disorder. Previous literature has suggested that the effects of growing up with a father with AUD can manifest differently than those who had grown up with an affected mother or two affected parents (e.g., \cite{steinhausen1984psychopathology,stout1996differences}), so we focus our investigation on the long-term impacts of growing up with a father who had an alcohol use disorder. Our findings will supplement and extend previous results on this topic, such as those recorded by \cite{rydelius1981children} and \cite{thor2022fathers}.

We will examine the impacts of growing up with a father with AUD on long-term economic success, interpersonal relationships, physical, and mental health using data from the Wisconsin Longitudinal Study (WLS), a long-term study of a random sample of over ten thousand individuals who graduated from Wisconsin high schools in 1957 (see, for instance, \cite{sewell2003we,herd2014cohort}). These data contain detailed information on the participants that were collected at various times over the decades since their graduation, making it uniquely suited to facilitate the analysis of later life outcomes as graduates reach their 50s, 60s, and 70s. In the 2011 wave of data collection, WLS respondents were asked, ``When you were growing up, that is during your first 18 years, did you live with anyone who was a problem drinker or alcoholic?'' Among the over one thousand graduates who answered affirmatively, more than 750 cited growing up with an affected father. This large sample size will help facilitate powerful data analysis and may lead to new discoveries that extend the current literature. The WLS also collected detailed measurements of variables that may confound the relationship between growing up with an affected parent and later life outcomes, such as those related to personal demographics and family background. We will use these variables to create matched sets of treated and control graduates, allowing us to compare the outcomes of similar individuals and lessen the burden of confounding biases. 

We introduce a novel statistical design, called \textit{data turnover}, to conduct this analysis. Data turnover allows for a single group of statisticians and domain experts to work together to assess the strength of evidence gathered across multiple data splits, while incorporating both qualitative and quantitative findings from data exploration. Exploratory data analysis allows us look at our data before making any (potentially dubious) assumptions, while enabling the recognition of any obvious errors, detection of outliers or anomalies, and shaping of new hypotheses. It also permits us to identify a variable that does not measure what we initially thought and help suggest a new one, or facilitate data-informed hypothesis selection (\cite{bekerman2024planning}). Along with exploration, replication is a valuable tool for statisticians conducting observational studies. Replication of scientific results lends credence to these findings and serves to maintain and build trust in scientific knowledge as a whole. Previous works have facilitated analyses that allow for exploration but no replication (e.g., \cite{cox1975note,heller2009split}), replication but no exploration (e.g., \cite{zhao2018cross, karmakar2019integrating}), or both under the stringent requirement of two independent teams of investigators (\cite{roy2022protocol}). Data turnover allows us to explore the data and assess the replicability of our findings with only one group of researchers.

Evaluating the replicability of our findings will be made possible by analyzing the data separately on two discrete subpopulations of WLS graduates -- those whose father graduated high school and those whose father did not. In observational studies like ours where we do not control assignment of individuals to receive treatment, obtaining similar results across multiple groups with different treatment assignment mechanisms strengthens the evidence that the treatment is in fact the cause of its ostensible effects (\cite{rosenbaum2001replicating,rosenbaum2015see}). Paul Rosenbaum (\citeyear{rosenbaum2015cochran}) would liken our approach to accumulating evidence in a crossword puzzle: while each piece of evidence may be tentative when considered individually, assessing them jointly can lead to more convincing conclusions. Notably, the reasons for a more educated father drinking heavily may differ from those of a less educated father. For instance, those with less education might be more likely to live in neighborhoods with many alcohol retailers or more relaxed community norms with respect to excessive drinking (\citep{crum1993level}). Meanwhile, an individual who graduated high school and continued his studies might have been exposed to excessive drinking at university and developed an unhealthy relationship with alcohol. Despite increased prevention efforts, binge drinking remains frequent among college students and one study found that roughly 40\% of college students have binged in the previous two weeks (\cite{wechsler2002trends,clapp2003failed}). Employing data turnover, we can pre-specify how one group will be used to plan the analysis on the other group, and use exploratory data analysis on the other group to determine how to conduct inference on the original subpopulation. We refer to those outcomes for which the effect is significant on both groups as ``replicable'' findings (\cite{bogomolov2023replicability}). In Section \ref{analysis}, we provide a detailed description on how we use this non-random cross-screening approach to identify replicable outcomes. We also discuss testing of the global null hypothesis; that is, identifying those outcomes which are affected on at least one of the two subpopulations.

The remainder of the document is organized as follows. We introduce the WLS dataset and describe our data processing in Section \ref{wls}. In Section \ref{sec:matching}, we outline our matching methodology. We report our primary outcomes of interest in Section \ref{sec:outcomes}. Finally, we conclude with a description of our data analysis procedure in Section \ref{analysis}. %

\section{Wisconsin Longitudinal Study (WLS) Dataset} \label{wls}

The WLS is an extensive, long-term study of 10,317 individuals who graduated from Wisconsin high schools in 1957. Detailed information pertaining to family background, childhood experiences, education, health, interpersonal relationships, economic status, and much more, were collected from the graduates or their parents across multiple waves from  1957 to 2020. These data are broadly representative of white American men and women, typically of German descent, who have graduated high school. The WLS sample does not well-represent certain large strata of American society, such as those of minority status and those without at least a high school education.

To investigate the myriad long-term impacts of growing up with a father with AUD, we examine outcomes collected during the 1992/1993 wave of data collection, where most graduates were in their early to mid 50s. We restrict our sample to those who completed both the phone interview and mail questionnaire to help mitigate issues with missingness due to non-response or censoring by death. We also omit those individuals who revealed growing up with more than one individual with AUD, those who reported an individual who was not their father, and subjects who did not report their father's education level. Since we also have access to substantial information from siblings of the graduates, we impute any missing family background or treatment status data using answers provided by their siblings, if these data are available. Among the 2817 remaining graduates whose father did not complete a high school education, 2382 did not grow up with a problem drinker in the household, while 435 grew up with a father with AUD. On the other hand, there are 1846 remaining graduates whose father graduated high school, among which 1623 did not grow up with an affected father and 223 grew up with an affected father.

\section{Matching} \label{sec:matching}

In order to lessen the burden of potential confounding biases, we create matched sets of treated and control graduates based on relevant baseline covariates to identify alike individuals and compare their outcomes. We employ matching separately for participants whose fathers did not graduate high school and those whose fathers completed their high school education. We list these variables on which we match in Appendix \ref{covariates-list}. 

It would be preferable to match treated and control graduates with exactly the same baseline covariates, yet doing so is difficult or impossible to implement when considering several covariates. Consequently, we instead look to create matched sets such that the mean value of each covariate among the matched treated subjects is close to that among the matched control subjects. We refer to this as covariate balance, which we assess using the absolute raw mean difference between matched treated and control units for binary covariates, and absolute standardized mean difference for continuous variables.

To deal with missingness in our baseline covariates, we assume that, conditional on the observed covariate values and pattern of missingness, the probability of treatment does not depend on the missing covariate values (\cite{rosenbaum1984reducing}). We can then append indicator variables for the pattern of missing data and impute any missing values for each covariate using the mean of the non-missing values. In each subpopulation of graduates, we remove subjects who lack overlap based on the pooled within group standard deviation of the estimated logit propensity score (\cite{hansen2009propensity}). We then match each treated graduate to three controls via propensity score matching with a caliper, while exactly matching on sex, to achieve adequate covariate balance. Including multiple control units in each matched set allows us to improve the statistical efficiency of our estimators (\cite{rosenbaum2010design}, Chapter 8).

Among those graduates whose father did not complete a high school education, we obtain 434 matched sets (effective sample size 651), while we get 223 matched sets (effective sample size 334.5) of individuals whose father graduated high school. Effective sample sizes are calculated as the sum of the harmonic means of the number of units in treatment and control for each matched group
(\cite{hansen2011propensity}). Our matching successfully balances the individual covariates between treated and control groups, which we display graphically using Love plots in Appendix \ref{covariates-balance}.

\section{Pre-Specified Outcomes of Interest} \label{sec:outcomes}

We collect representative outcomes related to health, well-being, interpersonal relationships, substance use, and economic success to examine the diverse long-term effects of growing up with a father with AUD. We list the corresponding variable name(s) in the WLS dataset, gathered during the 1992/1993 wave of data collection, in parentheses. \vspace{1em}
\newline
\textbf{Physical Health and Wellness}
\begin{itemize}   
    \item {Personal health rating. (\texttt{z\_mx001rer})}
\end{itemize}
\textbf{Mental Health and Wellness}
\begin{itemize}
    \item {Depression via modified CES-D. (\texttt{z\_mu001rec})}
\end{itemize}
\textbf{Interpersonal Relationships}
\begin{itemize}
    \item {Times meeting family + friends last month. (\texttt{z\_mz023rer, z\_mz024rer})}
\end{itemize}
\textbf{Substance Use}
\begin{itemize}
    \item Your alcohol use ever affect work or home? (\texttt{z\_ru032re, z\_ru033re})
\end{itemize}
\textbf{Economic Status}
\begin{itemize}
    \item {Total income in the last 12 months. (\texttt{rp015ree})}
\end{itemize}

\pagebreak

\section{Design of Data Analysis} \label{analysis}

We seek to identify which of the aforementioned outcomes were affected by growing up with a father affected by AUD. We will deem those outcomes for which the effect of the treatment is found to be significantly negative or positive in both subgroups - those raised by a more educated father and those with a less educated father - as \textit{replicable} findings. We will also establish which of the outcomes are affected by the treatment on at least one of the two subpopulations, a weaker yet still informative standard of evidence akin to meta-analysis.

Our inference will be based on matched sets of graduates in the WLS sample, as described previously in Section \ref{sec:matching}. We introduce the following notation: for each outcome $k =1, \dots, 5$, we observe $I_k^{HS}$ matched differences among those whose father graduated high school, and $I_k^{NHS}$ matched differences among those who did not. We use subscript $k$ since each outcome may have a different number of missing values, so the number of matched pairs may vary across outcomes (see Table \ref{tab:missingness} in Appendix \ref{descrip-sets}). For each outcome $k$, we  consider matched sets $i=1,\dots,I_k^{HS}$ of graduates with more educated fathers, where each matched set has four subjects, one of which is treated with $Z_{ij}^{HS} = 1$, while the others are controls with $Z_{ij}^{HS} = 0$, such that $\sum_{j=1}^{4} Z_{ij}^{HS} = 1$. Sets have been matched for observed covariates such that $x_{ij} \approx x_{ij'}$ for all $i, j, j'$. Let $r_{T_{ijk}}^{HS}$ and $r_{C_{ijk}}^{HS}$ denote the treated and control potential outcomes for the $j$th subject in the $i$th matched set when assigned to treatment and control, respectively, for $k=1,\ldots, 5$. The observed value of outcome $k$ for this individual is $R_{ijk}^{HS} = Z_{ij}r_{T_{ijk}}^{HS} + (1 - Z_{ij}) r_{C_{ijk}}^{HS}$; thus, the individual treatment effect $\delta_{ijk}^{HS} = r_{T_{ijk}}^{HS} - r_{C_{ijk}}^{HS}$
is not calculable (\cite{neyman1923application, rubin1974estimating}). Let
$Z_k^{HS} = (Z_{11k}^{HS}, \dots, Z_{I_k^{HS} 4k}^{HS})$ be the vector of treatment assignments and $R_k^{HS} = (R_{11k}^{HS}, \dots, R_{I_k^{HS} 4k}^{HS})$
be the vector of observed responses for outcome $k$. These definitions are taken the same for individuals whose father did not graduate high school, for whom we replace superscript $HS$ by superscript $NHS$.

For each outcome $k$, Fisher’s sharp null hypothesis (\cite{fisher:1935}) for each subpopulation is
\begin{align*}
   H_{0k}^{HS} : \delta_{ijk}^{HS} = 0, \qquad i = 1, \dots, I_k^{HS}, \, j = 1,2,3,4.\\
   H_{0k}^{NHS} : \delta_{ijk}^{NHS} = 0, \qquad i = 1, \dots, I_k^{NHS}, \, j = 1,2,3,4.
\end{align*}

We assume Fisher's sharp null and employ randomization inference to compute one-sided $p$-values.
For each outcome $k$, we assess replicability by testing the composite null hypothesis given by the union of the sharp null hypotheses for both subpopulations of graduates, for which a rejection will allow us to conclude the existence of a treatment effect in each subgroup. We also test the global null by examining the intersection of the sharp null hypotheses for both subpopulations, for which a rejection will allow us to conclude the existence of a treatment effect in at least one subgroup. We propose a novel strategy to control the type I error rate of our analysis. Let $\ell(k)\in\{0,1,2\}$ be the number of subgroups where Fisher's sharp null hypothesis regarding outcome $k$ is false. Our procedure outputs for each outcome $k=1,\dots,5$ a lower bound $\hat{\ell}(k)$ on the number of subgroups where Fisher's sharp null hypothesis regarding this outcome is non-null. A lower bound $\hat{\ell}=0$ is a trivial lower bound, a lower bound $\hat{\ell}=1$ corresponds to the claim that outcome $k$ is non-null in at least one of the parts, i.e., a claim that the global null is false for this outcome. A lower bound $\hat{\ell}=2$ corresponds to the claim that outcome $k$ is non-null in both parts, i.e., a replicability claim. Our \textit{data turnover} method will guarantee control of the family-wise error rate (FWER) for these lower bounds, i.e., the probability that for at least one outcome $k,$ its lower bound $\hat{\ell}(k)$ exceeds the true number of subgroups where Fisher's null for this outcome is false, at no more than $\alpha = 0.05$ (see \cite{benjamini2008screening} and \cite{benjamini2009selective} for similar error measures). Our new method may increase detection power compared to automated cross-screening (\cite{zhao2018cross}) by allowing for a data-informed design plan and facilitating qualitative and quantitative insights from exploration, as well as the potential formation of novel hypotheses.

\subsection{Data Turnover}

\begin{steps}
      \item [\textbf{Step 1:}] Use the data for those graduates whose father did not graduate high school, hereafter referred to as subpopulation 1, to select the direction of the alternative regarding each outcome for those whose father did graduate high school, hereafter referred to as subpopulation 2, and select the hypotheses which will be tested for this latter subgroup. Finally, test the selected hypotheses for subpopulation 2.
    \begin{itemize}
        \item[-] \textit{Selection of directions of alternative for subpopulation 2:} For each non-binary outcome $k = 1,2,3, 5$, test on subpopulation 1 using a weighted version of $M$-statistics, with weights of $(m,\underline{m},\overline{m}) = (20,12,20)$ as proposed by \cite{rosenbaum2014weighted}, and return the right-sided and left-sided $p$-values, $p_k^{NHS}$ and $q_k^{NHS}$, respectively. For binary outcome $k=4$, follow the same procedure, but instead conduct McNemar's test to generate the $p$-values. Select the right-sided alternative for outcome $k$ for subpopulation 2 if $p_k^{NHS} < q_k^{NHS}$, and select the left-sided alternative otherwise. Define $D_k^{HS} = 1$ if right-sided alternative was selected and $D_k^{HS} = -1$ otherwise.
    \item[-] \textit{Selection of hypotheses for subpopulation 2:} For each outcome $k = 1,\dots, 5$, let $\Tilde{p}_k^{NHS} = \min(p_k^{NHS}, q_k^{NHS})$ be the minimum of the two one-sided $p$-values for subpopulation 1. Select the hypotheses to be tested for subpopulation 2 based on each $\Tilde{p}_k^{NHS}$ as follows: the hypothesis regarding outcome $k$ is selected if $\Tilde{p}_k^{NHS} < \alpha/2$. Let $\mathcal{H}^{{HS}} \subseteq \{1,\dots, 5\}$ be the set of indices of selected hypotheses, i.e., $\mathcal{H}^{{HS}} = \{k: \Tilde{p}_k^{NHS} < \alpha/2\}.$ If there are no outcomes for which $\Tilde{p}_k^{NHS} < \alpha/2$, take $\mathcal{H}^{{HS}}$ to be the outcome $k$ which has the minimum $p$-value across all outcomes.
    \item[-] \textit{Testing the hypotheses for subpopulation 2:} For each selected hypothesis with index $k \in \mathcal{H}^{{HS}}$, compute the one-sided $p$-value $p_k^{HS}\1(D_k^{HS} = 1) + q_k^{HS}\1(D_k^{HS} = -1),$ where $p_k^{HS}$ and $q_k^{HS}$ are the right-sided and left-sided $p$-values for outcome $k$ computed using the data for subpopulation 2, and $\1(\cdot)$ is the indicator. Apply Holm's procedure (\cite{holm1979simple}) on $\{p_k^{HS}\1(D_k^{HS} = 1) + q_k^{HS}\1(D_k^{HS} = -1), k \in \mathcal{H}^{{HS}}\}$ at level $\alpha/2$, where $\alpha$ is the target FWER level for lower bounds. Let $\mathcal{S}^{HS} \subseteq \mathcal{H}^{{HS}}$ be the set of indices of outcomes such that for each $k \in \mathcal{S}^{HS}$, the null hypothesis for outcome $k$ was rejected on subpopulation 2.
    \end{itemize}
\item [\textbf{Step 2:}] In this step, we are able to explore the data in subpopulation 2 and choose in any manner how to conduct one-sided hypothesis tests for subpopulation 1. Test these hypotheses for subpopulation 1 and return the subset of rejected hypotheses, $\mathcal{S}^{NHS}\subseteq \{1, \ldots, 5+m\},$ where $m\geq 0$ is the number of any novel (i.e., not pre-specified) hypotheses tested for subpopulation 1, along with $D_k^{NHS},$ the indicator for the direction of the alternative for each $k\in \mathcal{S}^{NHS},$ where $D_k^{NHS}=1$ for right-sided alternative and $D_k^{NHS}=-1$ for left-sided alternative.
\item [\textbf{Step 3:}] For each $k \in \{1, \dots, 5+m\}$, set $\hat{\ell}(k)= \1(k\in \mathcal{S}^{NHS}) + \1(k\in \mathcal{S}^{HS}),$ where $\hat{\ell}(k)$ is the lower bound for the number of subpopulations for which the null hypothesis regarding outcome $k$ is false. Declare outcome $k$ as replicable if and only if $\hat{\ell}(k) = 2$ and $D_k^{NHS} = D_k^{HS}$. Reject the global null for outcome $k\in\{1, \ldots,5\}$ or a novel hypothesis with index $k\in\{6, \ldots, m\}$ if and only if $\hat{\ell}(k) \geq 1$.
\end{steps}

One intriguing question is, when working with split samples of unequal size, would it be more advantageous to pre-specify our analysis plan on the smaller or larger sample? Preliminary simulation results suggest the latter strategy, so that is how we describe our analysis plan above. We expect that incorporating data exploration in one subpopulation (see Step 2) will enhance the overall design of our study by facilitating data-informed decisions into our design and accommodating the possible development of new hypotheses. The flexibility afforded by our approach may allow for even further benefits we have not outlined nor yet anticipated. For replicability analysis, we are restricted to the pre-defined hypotheses in Section \ref{sec:outcomes}, but this constraint is removed for the analysis of the global null hypotheses and when conducting inference on each subgroup separately. This flexibility allows us to consider more complex null hypotheses, for instance, examining whether there is any effect modification among the outcomes. Moreover, the WLS dataset is richer than the list of outcomes outlined in Section \ref{sec:outcomes}, so an unrestricted look at the data may facilitate the development of additional novel hypotheses on related outcomes. For example, if self-rated physical health and wellness was affected, we might be interested in assessing which maladies in particular may have been driven by growing up with a father with AUD.

Moreover, we are also interested in reporting effect sizes for the rejected null hypotheses with well-defined parameters. Along with forming marginal confidence intervals for the selected parameters, we also suggest constructing on each subgroup $\alpha/2$ level false coverage rate (FCR; \cite{benjamini2005false}) confidence intervals for these parameters, thereby providing an overall $\alpha$-level FCR guarantee on all confidence intervals constructed. On each subpopulation, we will determine the parameters for which a confidence interval will be constructed if the hypotheses regarding these parameters are rejected. If for a certain subpopulation where $h$ hypotheses were tested, $r$ parameters were selected at level $\alpha/2$, we will construct the confidence intervals for these parameters at level
$1-r\alpha/(2h).$ Consequently, the expected fraction of non-covering confidence intervals (i.e., the FCR) for each subgroup will be bounded above by $\alpha/2$.

\bibliography{../references.bib}

\orignewpage
\appendix

\section{Baseline Covariates} \label{covariates-list}

We gather variables related to personal demographics and family background to create matched sets of treated and control graduates. We list the corresponding variable name in the WLS dataset in parentheses, adding a $^*$ if the covariate was constructed (i.e., modified) from the listed variable. These covariates were gathered over two waves delineated below:

-- Collected in 1957:
\begin{itemize}
    \item Coarsened population of town in which graduate attended high school. (\texttt{rlur57$^*$})
    \item Year of birth. (\texttt{z\_brdxdy})
    \item Sex of respondent. (\texttt{z\_sexrsp})
\end{itemize}

-- Collected in 1975:
\begin{itemize}
    \item Father's years of education in 1957. (\texttt{bmfaedu})
    \item Father's Duncan SEI score for 1957 job. (\texttt{bmfoc3u})
    \item Mother's years of education in 1957. (\texttt{bmmaedu})
    \item Natural log of parental income in 1957. (\texttt{bmpin1$^*$})
    \item Coarsened nationality of father. (\texttt{natfth$^*$})
    \item Coarsened religious preference of family in 1957. (\texttt{relfml$^*$})
    \item Total number of siblings. (\texttt{sibstt})
\end{itemize}

\orignewpage

\section{Covariate Balance} \label{covariates-balance}

\begin{figure}[b]
\centering
\includegraphics[width = \textwidth-6em]{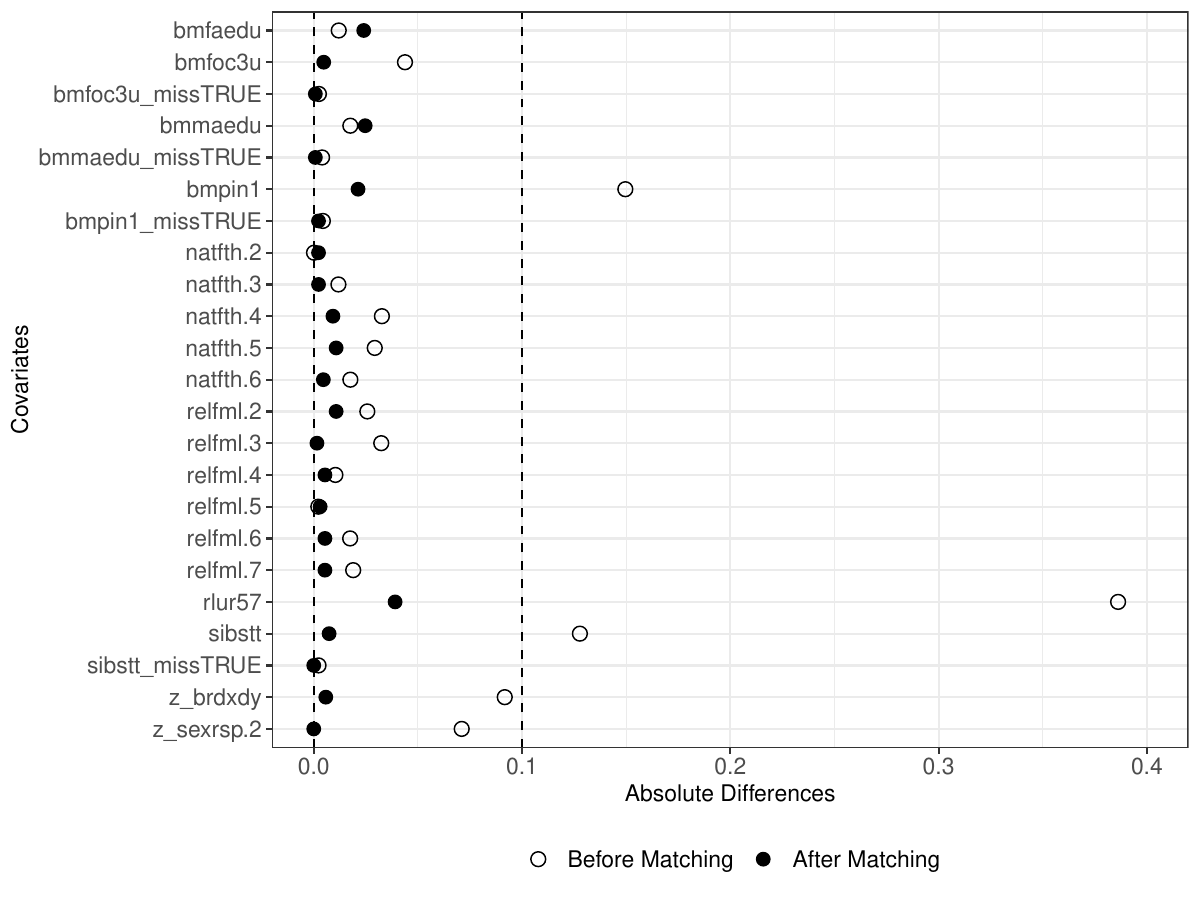}
\caption{Love plot showing covariate balance before and after matching among those whose fathers are not high school graduates. Absolute standardized mean differences are used to measure continuous covariates, while absolute mean differences are used for binary covariates.}
\label{fig:noHS}
\end{figure} 

\begin{figure}[]
\centering
\includegraphics[width = \textwidth-6em]{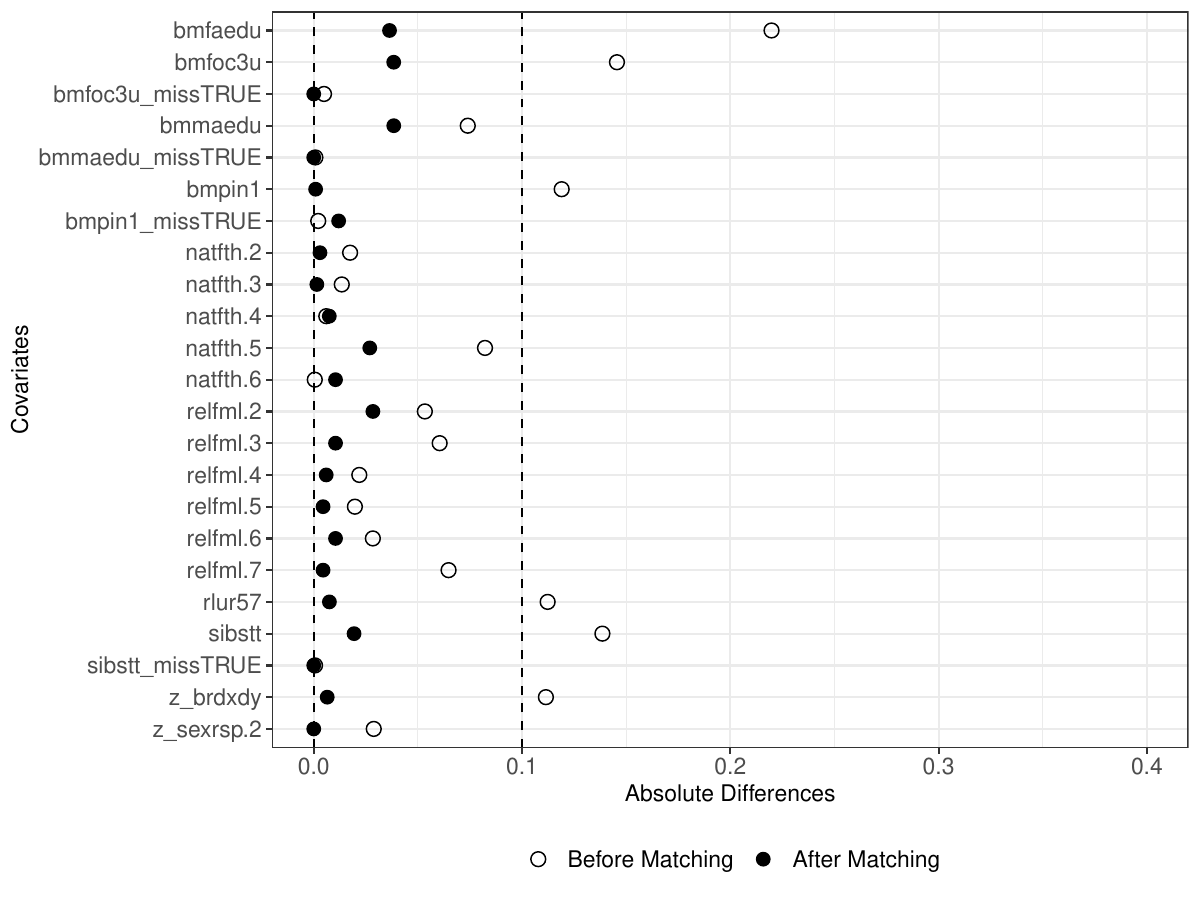}
\caption{Love plot showing covariate balance before and after matching among those whose fathers are high school graduates. Absolute standardized mean differences are used to measure continuous covariates, while absolute mean differences are used for binary covariates.}
\label{fig:HS}
\end{figure}

\orignewpage

\section{Description of Matched Sets} \label{descrip-sets}

\begin{table}[htbp]
  \centering
  \caption{Number of matched sets with non-missing values for each pre-specified outcome of interest.}
  \label{tab:missingness}
  \begin{threeparttable}
    \begin{tabular}{@{} lcc @{}}
      \toprule
      \multirow{2}{*}{\textbf{Outcome}} & \textbf{NHS (Subpop. 1)} & \textbf{HS (Subpop. 2)} \\
      \cmidrule(lr){2-2} \cmidrule(lr){3-3}
      & Number of Sets & Number of Sets \\
      \midrule
      Personal Health Rating & 432 & 222 \\
      Depression Score & 421 & 216 \\
      Times Meeting Family \& Friends & 429 & 215 \\ 
      Alcohol Affect Work or Home & 161 & 86 \\
      Income & 431 & 223 \\
      \bottomrule
    \end{tabular}
    \begin{tablenotes}
      \footnotesize
      \item \textit{Note.} NHS = graduates whose father did not graduate high school. \\HS = graduates whose father completed high school education.
    \end{tablenotes}
  \end{threeparttable}
\end{table}

\end{document}